\documentstyle[pra,aps,epsf,twocolumn]{revtex}  % DON'T CHANGE
\begin{document}                % INITIALIZE - DONT CHANGe
\title{Spontaneous emission and level shifts in absorbing 
disordered dielectrics and dense atomic gases:
 A Green's function approach}
\author{Michael Fleischhauer}
\address{Sektion Physik, Ludwig-Maximilians Universit\"at M\"unchen,
D-80333 M\"unchen, Germany}
\date{February 11. 1999}
\maketitle 
%\tighten

%%%%%%%%%%%%%%%%%%%%%%%%%%%%%%%%%%%%%%%%%%%%%%%%%%%%%%%%%%%%%%%%%%%%%%%%%%%%%%

\begin{abstract} 
Spontaneous emission and Lamb shift of atoms in absorbing dielectrics
 are discussed. A Green's-function approach is used 
based on the  multipolar interaction Hamiltonian 
of a collection of atomic dipoles with the quantised radiation field. 
The rate of decay and level shifts
are determined by the retarded Green's-function of the 
interacting electric displacement field, which is
calculated from a Dyson equation describing multiple
scattering. 
The positions of the atomic dipoles forming the dielectrics
are assumed to be
uncorrelated and a continuum approximation is used. 
The associated unphysical interactions 
between different atoms at the same location is eliminated by
removing the point-interaction term from the free-space Green's-function
(local field correction). For the case of an atom in a purely dispersive
medium the spontaneous emission rate is altered by the well-known 
Lorentz local-field factor. In the presence of absorption a result 
different from previously suggested expressions is found and 
nearest-neighbour interactions are shown to be important.
\end{abstract}

%%%%%%%%%%%%%%%%%%%%%%%%%%%%%%%%%%%%%%%%%%%%%%%%%%%%%%%%%%%%%%%%%%%%%%%%%%%%%%


%\narrowtext

%\widetext

%%%%%%%%%%%%%%%%%%%%%%%%%%%%%%%%%%%%%%%%%%%%%%%%%%%%%%%%%%%%%%%%%%%%%%%%%%%%%%
%%%%%%%%%%%%%%%%%%%%%%%%%%%%%%%%%%%%%%%%%%%%%%%%%%%%%%%%%%%%%%%%%%%%%%%%%%%%%%

\section{Introduction}

%%%%%%%%%%%%%%%%%%%%%%%%%%%%%%%%%%%%%%%%%%%%%%%%%%%%%%%%%%%%%%%%%%%%%%%%%%%%%%
%%%%%%%%%%%%%%%%%%%%%%%%%%%%%%%%%%%%%%%%%%%%%%%%%%%%%%%%%%%%%%%%%%%%%%%%%%%%%%
The theoretical description and experimental investigation of
the interaction of light with dense atomic media regained considerable
interest in recent years. 
Various experiments on level shifts \cite{Woerdmann96,Woerdmann99},
intrinsic bistability \cite{Boyd91,Rand94} and spontaneous emission
\cite{Rikken95,Schuurmans98} in dense gases have supported and
refined the concept of local fields known for more than a century \cite{LL}. 
Nevertheless some questions in this context are still not answered 
satisfactory even on a fundamental level. In the present paper I want to
discuss one of these questions, namely the effect of an {\it absorbing} 
dielectric on  spontaneous emission and level shifts of an
embedded atom using a Green's-function approach.

The interaction of light with dilute gases is usually well described in terms 
of macroscopic classical variables such as electric field and polarisation. 
In the macroscopic approach 
the polarisation is given by the expectation value of 
the single-atom dipole moment 
multiplied by the density of atoms \cite{SSL}. Apart from the coupling to the
common classical radiation field, the atoms are assumed distinguishable 
and independent. This means quantum-statistical correlations
are neglected, which is a very good approximation as long as the
temperatures are not too small. It is also implicitly assumed 
that vacuum fluctuations of the field affect the atoms only individually and
that the atom positions are independent of each other. 
The latter assumptions are however no longer valid in dense samples.

If the resonant absorption
length of some atomic transition  becomes comparable to the
medium dimension $d$, i.e.~for $N\lambda^2 d\sim 1$,
$N$ being the number density and $\lambda$ the resonant wavelength,
reabsorption and
multiple scattering of spontaneous photons and associated effects
like radiation trapping \cite{Holstein} or, if atomic excitation is 
present, amplified spontaneous emission 
need to be taken into account. If the atomic density is further increased, 
such that $N\lambda^3\sim 1$, one can no longer disregard the
fact that the independent-atom approximation allows for an
unphysical interaction of different atoms at the {\sl same} position and 
Lorentz-Lorenz local field corrections are needed \cite{LL}.

The modification of the rate of spontaneous emission $\Gamma$ 
by the local environment
was first noted by Purcell \cite{Purcell46}. Alterations of this rate
have been demonstrated experimentally 
near dielectric interfaces \cite{Drexhage74},  in quantum-well
structures \cite{Yamamoto91} and in cavities \cite{CQED}. 
Based on an analysis of the density of radiation states
Nienhuis and Alkemade predicted for an atom embedded 
in a homogeneous transparent dielectric 
with refractive index $n$ \cite{Nienhuis76}:
\begin{eqnarray}
\Gamma=\Gamma_0 n\label{gamma_N}
\end{eqnarray}
where $\Gamma_0$ is the free-space decay rate.
\begin{equation}
\Gamma_0= \frac{\wp^2 \omega_{ab}^3}{3\pi \hbar \epsilon_0 c^3},
\end{equation}
 $\wp$ being the electric dipole moment of the
transition with frequency $\omega_{ab}$.
The alteration of spontaneous emission by the index of refraction
leads to interesting potential applications as the suppression or
enhancement of decay in photonic band-gap materials \cite{band-gap}.
The approach of Ref.\cite{Nienhuis76} did neither 
take into account local-field corrections nor absorption however.

There has been a considerable amount of theoretical work on local-field 
corrections to spontaneous emission of an atom in {\it lossless} homogeneous 
dielectrics. Essentially all approaches assume a small cavity around the
radiating atom and the theoretical predictions depend substantially on the
details of this local-cavity model. Approaches based on Lorentz's
``virtual'' cavity \cite{Knoester89,Milonni95}
lead to
\begin{equation}
\Gamma_{\rm Lor}=\Gamma_0\, n\, \left(\frac{n^2+2}{3}\right)^2,
\label{gamma_KM}
\end{equation}
while those based on a real empty cavity  
\cite{Glauber91} predict
\begin{equation}
\Gamma_{\rm emp}=\Gamma_0\, n\, \left(\frac{3n^2}{2n^2+1}\right)^2.
\label{gamma_GL}
\end{equation}
For pure systems or impurities in disordered non-absorbing 
dielectrics Eq.(\ref{gamma_KM}) is believed to be correct. On the other hand 
recent experiments with Eu$^{3+}$ ions in organic ligand cages
verified the real-cavity expression Eq.(\ref{gamma_GL}) 
\cite{Rikken95,Schuurmans98}. 
An explanation for the different results was very recently given 
by de Vries and Lagendijk \cite{deVries98}. Applying a rigorous microscopic
scattering theory for impurities in non-absorbing dielectric 
cubic crystals, they showed that the local environment
determines whether Eq.(\ref{gamma_KM}) or (\ref{gamma_GL}) should be used.
For a substitutional impurity the empty-cavity result applies, while
for an interstitial impurity the virtual-cavity formula is correct. 
The latter also supports the  believe
that Eq.(\ref{gamma_KM}) is the correct one for disordered systems
like gases. 

While the effect of a transparent dielectric on spontaneous
emission is rather well studied, this it not the case for {\it absorbing}
media. A first step in this direction was made by   
Barnett, Huttner and Loudon
\cite{Barnett92}. Based on a discussion of the retarded Green's-function
in an absorbing  bulk dielectric they showed, that the index of 
refraction in (\ref{gamma_N}) is to be replaced by the
real part $n^\prime$ of the complex refractive index
$n=n^\prime+i  n^{\prime\prime}$. They also argued that the
square of the Lorentz-local field factor in (\ref{gamma_KM})
should be replaced by the absolute square, leading to
\begin{equation}
\Gamma=\Gamma_0\, n^\prime(\omega_{ab}) 
\left\vert\frac{n^2(\omega_{ab})+2}{3}\right\vert^2.\label{gamma_B}
\end{equation}
In order to derive this equation Barnett et al.~postulated
in  \cite{Barnett96} an operator equivalent of the  
Lorentz-Lorenz relation between the Maxwell and local field.
This assumption has however some conceptual problems. 
As pointed out very recently by Scheel et al.
\cite{Scheel98}, an operator Lorentz-Lorenz relation cannot hold,
since both quantities, the Maxwell field and the local field 
 have to fulfil the same commutation relations.

In a recent paper we have developed an 
approach that takes into account local-field corrections as well as 
multiple scattering and reabsorption
of spontaneous photons in modified single-atom Bloch equations
\cite{Fleischhauer99}. The modified Bloch equations provide a
way of including dense-medium effects in a macroscopic approach.
In the present paper  expressions for the spontaneous emission rate
and Lamb-shift of an atom in a dense {\it absorbing} dielectric or a 
gas of identical atoms are derived 
following the approach of \cite{Fleischhauer99}. 
The starting point is the multipolar-coupling Hamiltonian in dipole
approximation. 
The retarded Green's-function of the electric displacement field,
which determines decay rate and Lamb shift, is
calculated from a Dyson equation in self-consistent Hartree approximation.
As the atom positions are assumed to be independent from each other,
local-field corrections are needed to remove unphysical
interactions between atoms at zero distance. This is done in the present 
approach by an appropriate modification of the free-space Green's-functions
rather than by introducing a cavity. The rate of spontaneous
emission derived coincides with the virtual-cavity result (\ref{gamma_KM})
for a transparent dielectric, but differs from Eq.(\ref{gamma_B}) in the
case of absorption. It will be shown that in the presence of absorption 
near-field interactions with neighbouring atoms become very important, whose
correct description requires however a fully microscopic approach.

%%%%%%%%%%%%%%%%%%%%%%%%%%%%%%%%%%%%%%%%%%%%%%%%%%%%%%%%%%%%%%%%%%%%%%

\section{Radiative interactions in dense atomic media}

%%%%%%%%%%%%%%%%%%%%%%%%%%%%%%%%%%%%%%%%%%%%%%%%%%%%%%%%%%%%%%%%%%%%%%

The present analysis is based on a description of the atom-field
interaction in the dipole approximation using the multipolar Hamiltonian
in the radiation gauge  \cite{Milonni84}
\begin{equation}
{\hat H}_{\rm int}=-\frac{1}{\epsilon_0}\sum_j{\hat {\vec d}}_j\cdot 
{\hat{\vec D}}(\vec r_j).
\end{equation}
Here ${\hat d}_j$ is the dipole operator of an atom at position $\vec r_j$.
${\hat D}$ is the operator of the 
electric displacement with $\nabla \cdot{\hat {\vec D}}=0$.

It was shown in \cite{Fleischhauer99} that the effects of radiative atom-atom
interactions in a dense medium can be described in Markov approximation
with a nonlinear density-matrix equation
%\widetext
\begin{eqnarray}
\dot\rho &=& -\frac{i}{\hbar}\Bigl[H_0,\rho\Bigr]+
i\frac{\wp_\mu}{\hbar}\biggl[\sigma_\mu {\cal E}_{L\mu}^- +
\sigma_\mu^\dagger {\cal E}_{L\mu}^+,\rho\biggr]
\nonumber\\
&& - i h_{\mu\nu} \Bigl[\sigma_\nu^\dagger\sigma_\mu,\rho\Bigr]
- i h^c_{\mu\nu}\biggl[\Bigl[\sigma_\nu^\dagger,\sigma_\mu\Bigr]
,\rho\biggr]\nonumber\\
&&-\frac{\Gamma_{\mu\nu}}{2}
\biggl\{\sigma_\nu^\dagger 
\sigma_\mu \rho + \rho \sigma_\nu^\dagger \sigma_\mu
-2 \sigma_\mu\rho\sigma_\nu^\dagger  \biggr\}\nonumber\\
&&  - \frac{\Gamma^c_{\mu\nu}}{2}
\biggl\{ \Bigl[\sigma_\mu,\bigl[\sigma_\nu^\dagger,\rho\bigr]\Bigr]
+\Bigl[\sigma_\nu^\dagger,\bigl[\sigma_\mu,\rho\bigr]\Bigr]\biggr\}.
\end{eqnarray}
Here $\wp_\mu$ is the dipole matrix element for a polarisation direction
$\vec e_\mu$ and $\sigma_\mu,\sigma_\mu^\dagger$ are the corresponding 
atomic lowering and razing operators. The first term describes the free
atomic evolution and the second the interaction with some local classical
field ${\cal E}_L$. 
$h_{\mu\nu}$ and $\Gamma_{\mu\nu}$ are matrices, whose eigenvalues
yield Lamb shifts of excited states and spontaneous emission rates.
$\Gamma^c_{\mu\nu}$ and $h^c_{\mu\nu}$ describe  collective relaxation
rates  and light-shifts due to the incoherent background
radiation generated by absorption and reemission of spontaneous photons
(radiation trapping). 

It should be noted that the incoherent background radiation causes
a decay as well as an incoherent excitation with equal rate $\Gamma^c$. Thus 
$\Gamma^c$, which is proportional to the excitation of the host medium
\cite{Fleischhauer99},
describes {\it induced } mixing processes, while $\Gamma$ can be
interpreted as the rate of {\it spontaneous} decay. 
Similarly $h^c$ describes a light-shift, which for a two level
system is equal in strength and opposite in sign for the ground and
excited state. It is also proportional to the excitation of the host medium
and can thus be interpreted as {\it induced} light shift. In contrast
$h$ is a frequency shift of an excited state only and does not require
excitation of the host medium. 

The matrices $\Gamma_{\mu\nu}$ and $h_{\mu\nu}$ are 
given by
\cite{Fleischhauer99}
\begin{eqnarray}
\Gamma_{\mu\nu} &=& 2\frac{\wp_\mu\wp_\nu}{\hbar^2} \,
   {\rm Re}\Bigl[ {\rm D}_{\mu\nu} 
      (0,\omega_{ab})\Bigr],\\
h_{\mu\nu}\,  &=& \frac{\wp_\mu\wp_\nu}{\hbar^2} \,
   {\rm Im} \Bigl[ {\rm D}_{\mu\nu} 
      (0,\omega_{ab})\Bigr],
\end{eqnarray}
where $ {\rm D}_{\mu\nu}(\vec x,\omega)
\equiv \int_{-\infty}^\infty {\rm d}\tau \, D_{\mu\nu}(\vec x,\tau) 
{\rm e}^{i\omega\tau} $ is the Fourier-transform of the
retarded Green's-function (GF)  of the electric displacement field
defined here as
\begin{equation}
{\rm D}_{\mu\nu} (\vec x,\tau)
=\theta(\tau)\bigl\langle 0 \bigr\vert 
\Bigl[{\hat D}_\mu(\vec r_1,t_1),
{\hat D}_\nu(\vec r_2,t_2)\Bigr] 
\bigl\vert 0\bigr\rangle\, \epsilon_0^{-2},
\end{equation}
with $\vec x=\vec r_1-\vec r_2$ and $\tau=t_1-t_2$.
In the case of randomly oriented two-level atoms, one can replace
$\wp_\mu\to\wp$ and perform an orientation average yielding a single 
decay rate $\Gamma$ and a single excited-state level shift $h$.

The dense atomic medium affects the spontaneous emission of a single
probe atom due to multiple scattering of virtual photons. 
The scattering process can formally be described by a Dyson equation
for the exact retarded GF
\begin{equation}
{\bf D}(1,2) = {\bf D}^0(1,2) -
\int\!\!\!\int d3\,d4\, 
 {\bf D}^0(1,3)\, {\bf \Pi}(3,4)\, 
{\bf D}(4,2).
\label{Dyson}
\end{equation}
Here the  
integration is over $t$ from $-\infty$ to $+\infty$ and the whole sample
volume. ${\bf D}^0$ is the (dyadic) GF in free space and  
${\bf \Pi}$ is a formal (dyadic) self-energy. 
As shown in \cite{Fleischhauer99},
the self-energy can be described for randomly oriented two-level atoms 
in self-consistent Hartree approximation by 
\begin{eqnarray}
{\bf\Pi}(1,2)&=&\sum_j \frac{2}{3}\frac{\wp^2}{\hbar^2}
\theta(t_1-t_2)
\Bigl\langle \bigl[\sigma_j^\dagger(t_1),\sigma_j(t_2)\bigr]
\Bigr\rangle\nonumber\\
&&\times \delta(\vec r_1-\vec r_j)\delta(\vec r_2-\vec r_j)
\, {\bf 1}.
\end{eqnarray}
${\bf 1}$ is unity matrix and 
$\sigma=|b\rangle\langle a|$ is the atomic spin-flip operator
from the excited state $|a\rangle$ to the lower state $|b\rangle$
in the Heisenberg picture, i.e.~it contains all interactions.
The factor $2/3$ results from an orientation average.

We now make a continuum approximation and assume a homogeneous medium, 
such that
\begin{equation}
{\bf\Pi}(1,2) \longrightarrow p(t_1,t_2) 
 \delta(\vec r_1-\vec r_2)\, {\bf 1},
\end{equation}
where
\begin{equation}
p(t_1,t_2) = \frac{2}{3} \frac{\wp^2}{\hbar^2} N \theta(t_1-t_2)
{\overline{
\Bigl\langle \bigl[\sigma^\dagger(t_1),\sigma(t_2)\bigr]
\Bigr\rangle}}.
\end{equation}
The over-bar denotes an average over some possible inhomogeneous
distribution and $N$ is the number density of atoms.

With the above made approximations, the Dyson equation (\ref{Dyson})
contains also scattering processes between atoms at the same position.
In a continuum approximation
the probability of two point dipoles  being at the same
position is of measure zero. This  nevertheless leads
to a non-vanishing contribution, since the dipole-dipole interaction has
a $\delta$-type point interaction. 
This unphysical contribution needs to be removed by a
local-field corrections, which will be discussed in the following section.

%%%%%%%%%%%%%%%%%%%%%%%%%%%%%%%%%%%%%%%%%%%%%%%%%%%%%%%%%%%%%%%%%%%%

\section{local-field correction of free-space Green's-function 
          and Lorentz-Lorenz relation}

%%%%%%%%%%%%%%%%%%%%%%%%%%%%%%%%%%%%%%%%%%%%%%%%%%%%%%%%%%%%%%%%%%%%

The retarded Green's-function in free space ${\rm D}^0_{\mu\nu}(1,2)
=\theta(t_1-t_2)\bigl\langle 0 \bigr\vert 
\bigl[{\hat D}_\mu^0(1),{\hat D}_\nu^0(2)\bigr] 
\bigl\vert 0\bigr\rangle\epsilon_0^{-2}$, where $1;2; \dots$ stand for 
$\vec r_1, t_1; \vec r_2, t_2;\dots$ etc., is a solution of the 
homogeneous Maxwell equation
with $\delta$-like source term
\begin{eqnarray}
&&\left(\frac{1}{c^2}
\frac{\partial^2}{\partial t^2}+\nabla\times\nabla\times
\right)
{\bf D}^0(1,2)\nonumber\\
&&\qquad = -\frac{i\hbar}{\epsilon_0}\frac{\omega^2}{c^2}
\delta(\vec r_1-\vec r_2)\delta(t_1-t_2)\, {\bf 1}.
\end{eqnarray}
${\bf D}^0$ has a particularly simple form in reciprocal space
\cite{deVries98b}
\begin{eqnarray}
{{{\bf D}}}^0
(\vec q,\omega^+) &=&  \frac{i\hbar}{\epsilon_0}\frac{k^2}{
\left(k^2 + i 0\right){\bf 1} - q^2 {\bf\Delta}_q}\\
&=& \frac{i\hbar}{\epsilon_0}\left[
 \frac{k^2}{k^2-q^2 +i 0}{\bf \Delta}_q +
\frac{{\vec q}\circ{\vec q}}{q^2}\right],
\end{eqnarray}
where  $k=\omega/c$, and
${\bf \Delta}_q=
{\bf 1} - \frac{{\vec q}\circ{\vec q}}{q^2}$.
It may be worthwhile noting, that ${\bf D}^0$
is not transverse in $\vec q$-space, although 
$\nabla_1\cdot{\bf D}^0(1,2)\equiv 0$.
The corresponding function in coordinate space 
reads \cite{deVries98b}
\begin{eqnarray}
{ {\bf D}}^0(\vec x,\omega^+) &=&
-\frac{i\hbar\omega^2}{\epsilon_0c^2}
\frac{{\rm e}^{i k^+ x}}{4\pi x}\left[
P\left(i k x\right)\, {\bf 1} 
+Q\left(i k x\right)\frac{{\vec x}\circ{\vec x}}{x^2}
\right]\nonumber\\
&&  +\frac{i\hbar}{3\epsilon_0} \delta(\vec x)\, {\bf 1}.
\label{D0coord}
\end{eqnarray}
Here $x=|\vec x|$ and
\begin{eqnarray}
P(z) = 1-\frac{1}{z} +\frac{1}{z^2},\qquad
Q(z) = -1 + \frac{3}{z} -\frac{3}{z^2}.
\end{eqnarray}

One recognises from Eq.(\ref{D0coord}) that the retarded GF of the
dipole-dipole interaction contains a $\delta$-type point
contribution. In order to eliminate the unphysical interactions between
different atoms at the same position, one has to remove this 
term from the GFs in the scattering part of the Dyson equation
(\ref{Dyson}).
\begin{equation}
{{\bf D}}^0(\vec x,\omega^+) \longrightarrow 
{ {\bf F}}^0(\vec x,\omega^+)=
{{\bf D}}^0(\vec x,\omega^+) - \frac{i\hbar}{3\epsilon_0}
\delta(\vec x){\bf 1}.\label{local_correct}
\end{equation}
With this local-field corrections we obtain  a modified Dyson equation 
(in reciprocal space)
\begin{equation}
{\bf D}={\bf D}^0-{\bf F}^0 p {\bf F}^0 + 
{\bf F}^0 p {\bf F}^0 p {\bf F}^0  -+ \cdots,
\end{equation}
and introducing
 ${\bf F}(\vec q,\omega)\equiv{\bf D}(\vec q,\omega)-i\hbar/3\epsilon_0
\, {\bf 1}$
we  arrive at
\begin{equation}
{\bf F}(\vec q,\omega^+) = 
{\bf F}^0(\vec q,\omega^+) -
{\bf F}^0(\vec q,\omega^+)\, 
 p(\omega^+)\, 
{\bf F}(\vec q,\omega^+).
\label{Dyson2}
\end{equation}

In reciprocal space one finds 
\begin{eqnarray}
{{{\bf F}^0}}(\vec q,\omega^+)&=&
{{{\bf D}^0}}(\vec q,\omega^+)- 
\frac{i\hbar}{3\epsilon_0}{\bf 1},\\
&=&-\frac{i\hbar}{\epsilon_0}
\left[\frac{\left(\frac{1}{3}q^2 +\frac{2}{3}k^2\right){\bf 1} - 
{\vec q}\circ
{\vec q}}{q^2-k^2-i\epsilon}\right].
\end{eqnarray}
Eq.(\ref{Dyson2}) can easily be solved to yield
\begin{eqnarray}
{\bf F}(\vec q,\omega^+) &=& -\frac{i\hbar}{\epsilon_0}
\left[
\frac{\left(\frac{1}{3} q^2 +\frac{2}{3}k^2 \right)
\left(1+\frac{2}{3}N\alpha(\omega)\right){\bf 1} - {\vec q}\circ{\vec q}}
{q^2 -k^2 - N\alpha(\omega)\left(\frac{1}{3} q^2 +\frac{2}{3} k^2\right) 
-i0}\right]\nonumber\\
&&\quad\times \frac{1}{1+\frac{2}{3}N\alpha(\omega)}\label{Dyson_sol}
\end{eqnarray}
where we have introduced the dynamic polarisability of the atoms
\begin{equation}
N\alpha(\omega)\equiv \frac{i\hbar}{\epsilon_0} p(\omega).
\end{equation}
The poles $\pm q_0$ of Eq.(\ref{Dyson_sol}) determine the (in general 
nonlinear)
complex dielectric function
\begin{equation}
\varepsilon(\omega)\equiv\frac{q_0^2}{k^2} = 
1 +\frac{N\alpha(\omega)}{1-\frac{1}{3}N\alpha(\omega)}.\label{Lorentz-Lorenz}
\end{equation}
This is the well-known Lorentz-Lorenz relation between the microscopic
polarisability $\alpha$ and the complex dielectric function 
$\varepsilon(\omega)$.
Thus we have shown that the local-field correction of the free-space
Green's-function  (\ref{local_correct}) is exactly the one that
reproduces the well-known Lorentz-Lorenz relation.

%%%%%%%%%%%%%%%%%%%%%%%%%%%%%%%%%%%%%%%%%%%%%%%%%%%%%%%%%%%%%%%%%%%%

\section{Modification of spontaneous emission and Lamb-shift}

%%%%%%%%%%%%%%%%%%%%%%%%%%%%%%%%%%%%%%%%%%%%%%%%%%%%%%%%%%%%%%%%%%%%

Eq.(\ref{Dyson_sol}) can be transformed back into coordinate space
using ${\widetilde{\rm F}}(\vec x,\omega^+)=(2\pi)^{-3}\int {\rm d}^3\vec q 
\, {\widetilde{\rm F}}(\vec q,\omega^+)\, {\rm e}^{-i\vec q\vec x}$. 
The Fourier-transform of the projector $({\vec q}\circ{\vec q})$ yields
spherical Bessel functions \cite{deVries98b}. For the present purpose we
however need only the orientation-averaged quantity
\begin{eqnarray}
{\rm F}(\vec q,\omega^+)&=&
-\frac{2i\hbar}{3\epsilon_0}\left[\frac{ \frac{1}{3} q^2 N \alpha(\omega) 
+k^2\left(1+\frac{2}{3}N\alpha(\omega)\right)}
{q^2 -k^2 - N\alpha(\omega)\left(\frac{1}{3} q^2 +\frac{2}{3} k^2\right) 
-i0}\right]\nonumber\\
&&\quad\times\frac{1}{1+\frac{2}{3}N\alpha(\omega)}.
\end{eqnarray}

One recognises that the Fourier-transform of ${\rm F}(\vec q,\omega^+)$ 
diverges for $x\to 0$, which is due to the large-$q$ behaviour
of the GF. In order to remove these singularities one can modify
the GF by introducing a regularisation. Physically the
singular behaviour at $x\to 0$ is due to the fact that atoms very 
close to the atom
under consideration can have a large effect on spontaneous emission and
level shifts. One cannot expect the continuum approximation 
used here to yield accurate results on length scales comparable to the
mean atom distance. Here rather a fully microscopic description 
of very close atoms including their motion
(collisions) is needed. This is however beyond the scope of the present
paper and we therefore restrict the analysis to a regularisation of the
Green's-function. There is no unique regularisation procedure,
and we here just choose a convenient one
\begin{equation}
{\rm F}(\vec q,\omega^+)\longrightarrow
{\widetilde{\rm F}}(\vec q,\omega^+) = {\rm F}(\vec q,\omega^+) 
\frac{\Lambda^4}{q^4+\Lambda^{4}}.
\end{equation}
With this we find  in the limit $\Lambda\gg |q_0|$
\begin{eqnarray}
&&{\widetilde{\rm F}}(\vec x=0,\omega^+)
=\frac{\hbar \omega^3}{6\pi\epsilon_0c^3} \sqrt{\varepsilon(\omega)}
\left(\frac{\varepsilon(\omega)+2}{3}\right)^2\nonumber\\
&&\quad
 -\frac{i\hbar\omega^3}{6 \pi\epsilon_0 c^3}
\Biggl[\frac{1}{R} \left(\frac{\varepsilon(\omega)+2}{3}\right)^2\nonumber\\
&&\qquad\qquad\quad
+\frac{1}{R^3} \frac{2}{3}\left(\frac{\varepsilon(\omega)+2}{3}\right)
\bigl(\varepsilon(\omega)-1\bigr)\Biggr],
\end{eqnarray}
where $R=k/(\sqrt{2}\Lambda)$. 
It is important to note, that ${\widetilde{\rm F}}$ is exactly causal,
if $\varepsilon(\omega)$ fulfils the Krames-Kronig relations.
This would not have been the case if 
as according to the result of Barnett et al.\cite{Barnett92,Barnett96}
the absolute square $|(\varepsilon +2)/3|^2$ would be present
instead of $((\varepsilon +2)/3)^2$.

With this result we find for the decay rate and excited state
Lamb-shift
\begin{eqnarray}
\Gamma &=&\quad \Gamma_0\, {\rm Re}\left[ \sqrt{\varepsilon(\omega)} 
\left(\frac{\varepsilon(\omega)+2}{3}\right)^2\right]\nonumber\\
&&+\Gamma_0 \,
{\rm Im}\Biggl[\frac{1}{R} \left(\frac{\varepsilon(\omega)+2}{3}\right)^2
\nonumber\\
&&\qquad\qquad+\frac{1}{R^3} \frac{2}{3}\left(\frac{\varepsilon(\omega)+2}{3}\right)
\bigl(\varepsilon(\omega)-1\bigr)\Biggr],
\label{Gamma_result}\\
h &=& \quad \frac{\Gamma_0}{2}\, {\rm Im}\left[ \sqrt{\varepsilon(\omega)}
\left(\frac{\varepsilon(\omega)+2}{3}\right)^2\right]\nonumber\\
&& -\frac{\Gamma_0}{2}\, 
{\rm Re}\Biggl[\frac{1}{R} \left(\frac{\varepsilon(\omega)+2}{3}\right)^2
\nonumber\\
&&\qquad\qquad
+\frac{1}{R^3} \frac{2}{3}\left(\frac{\varepsilon(\omega)+2}{3}\right)
\bigl(\varepsilon(\omega)-1\bigr)\Biggr].\label{Lamb_result}
\end{eqnarray}
For an atom in a purely {\it dispersive} disordered medium, i.e.~for 
$\varepsilon^{\prime\prime}\equiv 0$, the second
term in Eq.(\ref{Gamma_result}) for the spontaneous decay
rate vanishes identically and we are left with the ``virtual'' cavity
result Eq.(\ref{gamma_KM}). Likewise there are no contributions from the
first term in (\ref{Lamb_result}) to the Lamb shift in this case.

In the presence of absorption, that is if the probe-atom transition
frequency comes closer to a resonance of the surrounding material
(as it would naturally be the case for a collection of identical atoms)
 $\Gamma$ is 
different from the result obtained in  
\cite{Barnett92,Barnett96,Juzeliunas97}.
In this case there are also non-vanishing terms that contain the
regularisation parameter $R^{-1}$ and $R^{-3}$. These terms
must be interpreted as contributions due to resonant energy transfer
with nearest neighbours, a process which cannot accurately
be described in the continuum approach used here.

As the Lamb shift is concerned, Eq.(\ref{Lamb_result}) shows that
in a purely dispersive medium, that is far away from any resonances
only nearest-neighbour interactions matter. This is intuitively clear
since in this case the transition frequency is only affected
by collisions. Only in the presence of absorption there is also
a bulk-contribution to the Lamb shift as described by the first term in
(\ref{Lamb_result}).

For a dense gas of identical atoms or of atoms of the same kind but 
with some inhomogeneous broadening, Eqs.(\ref{Gamma_result}) 
and (\ref{Lamb_result})
are only implicit, since the complex polarisability $\varepsilon$ 
depends on the decay rate and level shift. Hence a self-consistent 
determination  of $\Gamma$ and $h$ is necessary. If the density
of atoms is much less than one per cubic wavelength one can consider
an expansion of $\Gamma$ and $h$ in powers of the atomic density $N$.
Defining $\alpha=\alpha^\prime+i\alpha^{\prime\prime}$ one finds
with Eq.(\ref{Lorentz-Lorenz}) for the bulk contributions
\begin{eqnarray}
\Gamma &=& \Gamma_0\left[ 1+ \frac{7}{6}\alpha^\prime N +\frac{17}{24}
\left(\alpha^{\prime 2}-\alpha^{\prime\prime 2}\right) N^2 +{\cal O}(N^3)
\right],\\
h &=& \frac{\Gamma_0}{2}\left[\frac{7}{6}\alpha^{\prime\prime} N 
+\frac{17}{12}\alpha^\prime\alpha^{\prime\prime} N^2  +{\cal O}(N^3)\right].
\end{eqnarray}
In the case of radiatively broadened two-level atoms, the real part
of atomic polarisability vanishes at resonance, i.e. $\alpha^\prime=0$.
Thus in lowest order of the density there is only a contribution to the
excited state frequency proportional to the population difference between
excited and ground state. For an inverted population the transition
frequency is red-shifted, for balanced population the level shift
vanishes and for more atoms in the lower state the transition frequency
is blue shifted. As a result spontaneously emitted radiation 
from an initially inverted system will
have a chirp very similar to the chirp in Dicke-superradiance
\cite{Gross82}. It should also be mentioned that the
shift of the transition frequency discussed here is physically different
from the familiar Lorentz-Lorenz shift. The LL-shift is due to
the dispersion of the index of refraction at an atomic resonance
and is thus in contrast to the absorption $\alpha^{\prime\prime}$ 
independent on Doppler-broadening
\cite{Cooper96}.

%%%%%%%%%%%%%%%%%%%%%%%%%%%%%%%%%%%%%%%%%%%%%%%%%%%%%%%%%%%%%%%%%%%%%%%%%

\section{summary}

%%%%%%%%%%%%%%%%%%%%%%%%%%%%%%%%%%%%%%%%%%%%%%%%%%%%%%%%%%%%%%%%%%%%%%%%%%%

In the present paper we have discussed the rate of spontaneous emission
and the excited-state level shift of a two-level type 
probe atom inside a homogeneous, disordered
absorbing dielectric. The dielectric was modelled by a collection 
of atomic point dipoles,
which also includes the case of a dense gas of identical atoms. 
The multiple scattering of photons between the atoms (dipole-dipole
interaction) was described by a Dyson integral equation 
for the exact
retarded Green's-function of the electric displacement field
in self-consistent Hartree approximation.
The atoms were assumed distinguishable with
random independent positions. The latter assumption made a continuum
approximation possible and the Dyson equation could be solved analytically.
In order to exclude unphysical dipole-dipole interactions
of different atoms at the same position arising in the continuum approximation
with independent atomic positions, a local-field correction of the
free-space retarded Green's-function was introduced.
This lead to the well-known Lorentz-Lorenz relation between the
complex dielectric function $\varepsilon(\omega)$ and the
nonlinear atomic polarisability $\alpha(\omega)$. 
The expression for
the spontaneous-decay rate found by this method 
agrees with the virtual cavity result \cite{Knoester89}
in the absence of absorption. 
This is an expected result for atoms in disordered dielectrics
\cite{deVries98}. It was shown that the excited-state 
Lamb shift is in this case only affected by nearest-neighbour
interactions, which could not be treated accurately  within
the present approach however. 
In the presence of absorption the spontaneous-emission rate
differs from the results obtained in 
\cite{Barnett92,Barnett96,Juzeliunas97} in two ways. 
First there are important
nearest-neighbour contributions, which were absent in the
models of \cite{Barnett92,Barnett96}. Secondly the bulk-contribution
is different form Refs.\cite{Barnett92,Barnett96,Juzeliunas97},
since causality of the exact retarded GF requires the  
Lorentz-field factor to enter as square and not as absolute square.
It is interesting to note, that apart from a small difference 
in the $R^{-3}$ term, the decay rate derived here 
 is identical to one very recently obtained
by Scheel and Welsch \cite{Welsch99} on the basis of 
a completely different approach, namely a quantisation of 
the electromagnetic field in a linear dielectric.

%%%%%%%%%%%%%%%%%%%%%%%%%%%%%%%%%%%%%%%%%%%%%%%%%%%%%%%%%%%%%%%%%%%%%

\section*{Acknowledgement}
I would like to thank Charles Bowden, Janne Ruostekoski and Dirk-Gunnar Welsch
for stimulating discussions and 
D.G. Welsch and S. Scheel for making Ref.\cite{Welsch99} available prior
to publication.

%%%%%%%%%%%%%%%%%%%%%%%  begin references %%%%%%%%%%%%%%%%%%%%%%%%%%%%%%

\end{document}